\documentclass[aps,preprint]{revtex4}
\usepackage{graphicx}
\newcommand{\ket}[1] {\left| #1 \right\rangle}
\newcommand{\avg}[1] {\left\langle #1 \right\rangle}

\begin{document}

\title{Single-artificial-atom lasing using a voltage-biased superconducting charge qubit}

\

\author{S. Ashhab}
\affiliation{Frontier Research System, The Institute of Physical
and Chemical Research (RIKEN), Wako-shi, Saitama 351-0198, Japan}
\affiliation{Center for Theoretical Physics, Physics Department,
The University of Michigan, Ann Arbor, Michigan 48109-1040, USA}

\author{J. R. Johansson}
\affiliation{Frontier Research System, The Institute of Physical
and Chemical Research (RIKEN), Wako-shi, Saitama 351-0198, Japan}

\author{A. M. Zagoskin}
\affiliation{Frontier Research System, The Institute of Physical
and Chemical Research (RIKEN), Wako-shi, Saitama 351-0198, Japan}
\affiliation{Department of Physics, Loughborough University,
Loughborough LE11 3TU, United Kingdom}

\author{Franco Nori}
\affiliation{Frontier Research System, The Institute of Physical
and Chemical Research (RIKEN), Wako-shi, Saitama 351-0198, Japan}
\affiliation{Center for Theoretical Physics, Physics Department,
The University of Michigan, Ann Arbor, Michigan 48109-1040, USA}

\date{\today}

\begin{abstract}

We consider a system composed of a single artificial atom coupled
to a cavity mode. The artificial atom is biased such that the most
dominant relaxation process in the system takes the atom from its
ground state to its excited state, thus ensuring population
inversion. A recent experimental manifestation of this situation
was achieved using a voltage-biased superconducting charge qubit.
Even under the condition of `inverted relaxation', lasing action
can be suppressed if the `relaxation' rate is larger than a
certain threshold value. Using simple transition-rate arguments
and a semiclassical calculation, we derive analytic expressions
for the lasing suppression condition and the state of the cavity
in both the lasing and suppressed-lasing regimes. The results of
numerical calculations agree very well with the analytically
derived results. We start by analyzing a simplified two-level-atom
model, and we then analyze a three-level-atom model that should
describe accurately the recently realized superconducting
artificial-atom laser.

\end{abstract}


\maketitle

\section{Introduction}

Superconducting circuits have gained increased interest in recent
years, particularly for their possible use in quantum information
processing and as artificial atoms \cite{YouReview}. In relation
to the artificial-atom concept, the idea of placing such an atom
in contact with a harmonic-oscillator circuit element, which
serves as a cavity, has attracted a great deal of attention
\cite{Chiorescu}. Such circuit-QED systems hold promise for
studying various quantum-optics phenomena in a highly controllable
and easily tunable setting, as well as exploring parameter regimes
that are inaccessible using natural atoms.

One of the most intriguing and counterintuitive phenomena in the
fields of atomic physics and quantum optics is lasing
\cite{Milonni}. Given the above-mentioned advantages of
superconducting circuits for studying atomic-physics and
quantum-optics phenomena, it is natural to investigate
superconducting implementations of lasing. Indeed, there have been
a number of recent theoretical proposals
\cite{You,Rodrigues,Hauss,Zhirov,Hatakenaka} and experimental
demonstrations of lasing \cite{Astafiev,Grajcar} and population
inversion \cite{Berns} in superconducting systems.

In Ref.~\cite{You} a cyclically manipulated artificial atom is
constantly driven into its excited state, from which it can relax
by emitting a photon into the cavity, thus establishing a lasing
state. In Ref.~\cite{Hauss} an atom that is illuminated by an
oscillating field with a properly chosen frequency emits photons
into a low-frequency cavity. Here we analyze a situation that is
different from both Refs.~\cite{You,Hauss}, but is closer to the
usual picture of lasing with natural atoms. Furthermore, the
models that we study are closely related to the experiment of
Ref.~\cite{Astafiev}. It should be mentioned here that similar
models have been studied in the past in the study of single-atom
lasing (see e.g.~Refs.~\cite{Mu,Briegel}). A similar model was
also analyzed in Ref.~\cite{Rodrigues}, but that paper explored
different parameter regimes and analyzed different aspects of the
problem from the present paper.

Using a transition-rate-based calculation, a semiclassical
calculation and numerical simulations, we analyze the different
possible states of the cavity as the system parameters are varied.
Each one of the analytic calculations has its advantages. The
transition-rate-based calculation derives in a transparent manner
the lasing suppression condition and the state of the cavity deep
in the lasing and the suppressed-lasing regimes. The semiclassical
calculation provides a good approximation for the state of the
cavity throughout the lasing state, but is not suited to analyze
the suppressed-lasing regime, where it turns out that the state of
the cavity takes the form of a thermal state. For clarity, we
start by analyzing a simplified two-level-atom model, and we later
take the same approach to analyze a three-level-atom model that
describes more accurately the experiment of Ref.~\cite{Astafiev}.
In particular, we comment on a possible experimental
implementation of the crossover between the lasing and thermal
regimes with a superconducting artificial-atom laser.

\section{Two-level atom}

In this section we analyze the simplified model where the atom
contains two energy levels only. This model provides a good
qualitative understanding of the mechanisms at play and the
resulting phenomena in the experimental setup of interest to us.
The qualitative understanding developed in this section will also
be useful for identifying the importance of the different
processes in the more realistic model analyzed in Sec.~III below.

\subsection{Model}

\begin{figure}[h]
\includegraphics[width=7.0cm]{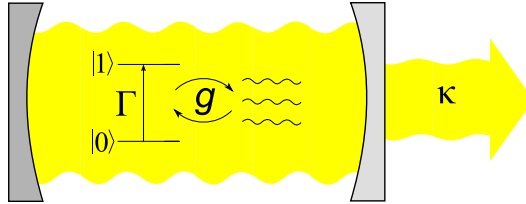}
\caption{(Color online) Schematic diagram of a two-level atom
interacting with a cavity mode. The coupling strength for the
exchange of excitations between the atom and the cavity is $g$.
The atom is biased such that it experiences `inverted relaxation'
from the ground to the excited state, with rate $\Gamma$. The loss
rate of photons out of the cavity is $\kappa$.}
\end{figure}

We consider the simple system composed of a two-level system
interacting with a harmonic oscillator (which typically is one
mode of an electromagnetic cavity). The system is shown
schematically in Fig.~1. The Hamiltonian of the combined
atom-cavity system is given by
\begin{equation}
\hat{H} = \frac{\hbar\omega_a}{2} \hat{\sigma}_z + \hbar\omega_0
\hat{a}^{\dagger} \hat{a} + g \sigma_x \left( \hat{a} +
\hat{a}^{\dagger} \right),
\end{equation}
where $\omega_a$ is the atom's characteristic frequency,
$\omega_0$ is the cavity's natural frequency, $g$ is the
atom-cavity coupling strength, $\hat{\sigma}_x$ and
$\hat{\sigma}_z$ are the usual Pauli matrices operating on the
atomic state, and $\hat{a}$ and $\hat{a}^{\dagger}$ are,
respectively, the annihilation and creation operators acting on
the state of the cavity. We shall describe quantum states using
the notation $\ket{n_a,n_c}$, where $n_a=0$ for the atomic ground
state and $n_a=1$ for the excited state, and $n_c$ represents the
number of photons in the cavity.

In order to have efficient emission of photons from the atom into
the cavity, the atom and cavity frequencies must be almost equal.
For the remainder of this paper, we shall take
$\omega_a=\omega_0$. We also take this frequency to be the largest
frequency (or energy) scale in the problem.

The setup is designed such that the atom's bias conditions cause
it to `relax' from the ground state to the excited state, with
rate $\Gamma$ \cite{2pi}. In this section, the inverted relaxation
is assumed as part of the theoretical model under consideration; a
similar process will be derived from first principles in a
realistic three-level-atom model in Sec.~III. It is this
counter-intuitive, inverted relaxation that will provide the
mechanism for population inversion, which plays a crucial role in
the realization of the lasing state. As such, one can say that the
(usual) threshold condition for lasing action is automatically
satisfied in this model. Note that we are ignoring any weak
relaxation process pushing the atom from the excited to the ground
state, since such a process would not affect the main points we
wish to study. Furthermore, since the atom's relaxation rate will
be taken to be very large, we shall ignore any additional atomic
dephasing mechanisms. The cavity is taken to possess a decay rate
$\kappa$.

An alternative description of the above situation concerning the
bias conditions would be to say that the cavity is in contact with
a heat bath that has a very small and positive temperature, while
the atom is in contact with a heat bath that has a very small and
negative temperature. It is worth mentioning here that a similar
approach (with negative effective bath temperature) was used in
Ref.~\cite{Gardiner} to describe an amplification process.

\subsection{Photon emission and loss rates}

In order to determine the state of the cavity for a given set of
parameters, we first note that the above model contains a
mechanism for photon emission into the cavity and a mechanism for
photon loss from the cavity. We consider these two mechanisms
separately.

The loss rate of photons from the cavity (i.e.~the transition rate
from the state $\ket{n_a,n}$ to the state $\ket{n_a,n-1}$, where
$n_a$ represents the state of the atom) is given simply by the
decay rate $\kappa$ multiplied by the number of photons in the
cavity $n$:
\begin{equation}
\Gamma_{\rm loss} = n \kappa.
\label{eq:LossRate}
\end{equation}

Obtaining the photon emission rate requires a somewhat more
careful analysis. We first consider the situation where there are
no or few photons in the cavity. The atom's bias conditions
constantly push it to its excited state. We can therefore assume
the atom to be initially in the excited state. If the atom is in
its excited state and the cavity has $n-1$ photons, the
atom-cavity coupling (with matrix element $g\sqrt{n}$) induces
dynamics between the states $\ket{1,n-1}$ and $\ket{0,n}$. Since
$\Gamma\gg g$, the dynamics will take the form of an incoherent
process described by the transition rate $W_{\ket{1,n-1}
\rightarrow \ket{0,n}} = 4n g^2/\Gamma$. Any population that
starts to accumulate in the state $\ket{0,n}$ will quickly relax
to the state $\ket{1,n}$, because the atom is constantly pushed in
this direction by its surrounding environment. These two steps
complete the transition from the state $\ket{1,n-1}$ to the state
$\ket{1,n}$, or in other words, the process of adding one photon
to the cavity. Since the upward-relaxation process occurs at a
very large rate, it can be treated as being instantaneous. We
therefore find that the photon emission rate (i.e.~the transition
rate from the state $\ket{n_a,n-1}$ to the state $\ket{n_a,n}$) is
given by the rate of the $\ket{1,n-1} \rightarrow \ket{0,n}$
transition, i.e.
\begin{equation}
\Gamma_{\rm emission} = \frac{4n g^2}{\Gamma}.
\label{eq:EmissionRate}
\end{equation}
The photon emission rate therefore increases linearly with $n$ for
small values of $n$ \cite{ExplainGamma1}. Clearly this situation
cannot persist for large $n$, since this mechanism is ultimately
limited by the atom's relaxation rate $\Gamma$. Indeed, when
$g\sqrt{n}$ becomes comparable to or larger than $\Gamma$, the
$\ket{1,n-1} \leftrightarrow \ket{0,n}$ transitions must be
treated as coherent oscillations. One can now argue that in the
limit of very large $n$, where the system spends half of the time
in each one of the two states ($\ket{1,n-1}$ and $\ket{0,n}$), the
atom has a chance to incoherently relax from its ground state into
its excited state only half of the time. In this case the photon
emission rate asymptotically reaches the value $\Gamma/2$, which
it cannot exceed.

The main advantage of the above derivation of the photon emission
rate is its simplicity, as well as the simplicity of the resulting
expressions. A more detailed analysis of the photon emission rate
for any value of $n$ is possible, assuming that cavity is in a
coherent (i.e.~lasing) state. This calculation will be carried out
in Sec.~II.D below (see also \cite{Mu,Stoof}).

\subsection{Lasing condition and possible steady states}

Combing the photon emission and loss rates as functions of photon
number $n$, one can obtain the probability distribution of photon
number states in the cavity. In particular, if this probability
distribution has a peak for some value of $n$, the peak value can
be obtained by locating the intersection point between the
emission and loss rates. Some relevant examples of such a
peak-finding calculation are depicted schematically in Fig.~2.

\begin{figure}[h]
\includegraphics[width=9.0cm]{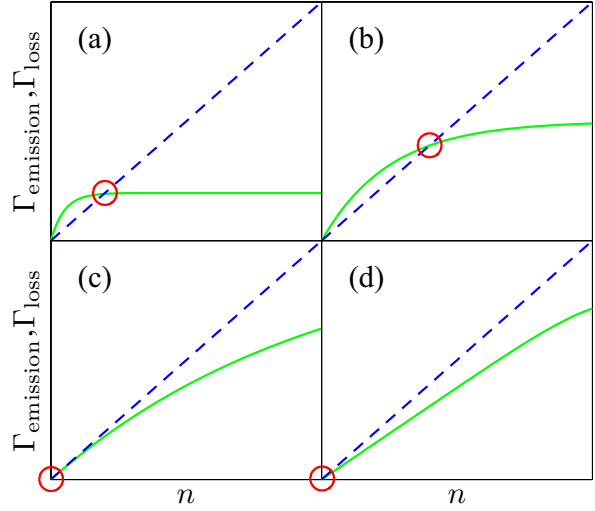}
\caption{(Color online) Schematic plot of the photon emission rate
$\Gamma_{\rm emission}$ (green solid line) and the photon loss
rate $\Gamma_{\rm loss}$ (blue dashed line) as functions of the
photon number in the cavity $n$. The intersection point (red
circle) determines the value of $n$ in the photon-number
probability distribution with the highest probability. Going from
(a) to (d), $\Gamma$ is increased, while $\kappa$ is kept fixed.}
\end{figure}

We treat the atom's relaxation rate $\Gamma$ as the tunable
parameter, keeping $g$ and $\kappa$ fixed. From
Eq.~(\ref{eq:EmissionRate}) one can see that small values of
$\Gamma$ correspond to a large initial slope of the emission rate
(at $n=0$), even though the emission rate reaches the saturation
level for a relatively small value of $n$. Figure 2(a) represents
this situation. If we increase $\Gamma$, the initial slope of the
emission rate decreases, but it eventually reaches a larger value.
By comparing Figs.~2(a) and 2(b), one can see that the peak value
of the photon number in the cavity increases with increasing
pumping rate. This result agrees with intuitive expectations.

A change of behaviour occurs when $\Gamma$ reaches a certain
regime, an expression for which will be given shortly. As can be
seen from Fig.~2(c), the peak value of $n$ starts decreasing with
increasing $\Gamma$ and vanishes at a certain value of $\Gamma$.
Beyond this point the value of $n$ with maximum occupation
probability remains at zero.

This suppression of lasing action by strong pumping is quite
counterintuitive. It can be understood in terms of the quantum
Zeno effect; at a certain point, the decoherence associated with
pumping becomes the most dominant effect and inhibits the emission
of photons from the atom into the cavity \cite{Wang,Gurvitz}. In
the following we shall analyze this effect, as well as the state
of the cavity in the different regimes, more quantitatively.

By combining Eqs.~(\ref{eq:LossRate}) and (\ref{eq:EmissionRate}),
we find that if
\begin{equation}
\frac{4 g^2}{\Gamma} > \kappa,
\label{eq:LasingCondition}
\end{equation}
the photon emission rate is larger than the photon loss rate,
assuming a small photon number in the cavity. Starting with a
small photon number, the number increases exponentially in time
[The growth in photon number continues until the peak value of $n$
is reached, as represented by the circles in Figs.~2(a,b)]. If, on
the other hand, Eq.~(\ref{eq:LasingCondition}) is not satisfied,
the loss rate will be higher than the emission rate, and lasing
would not occur. Equation (\ref{eq:LasingCondition}) can therefore
be considered a second threshold condition for lasing in this
setup. Note that population inversion is guaranteed in this model
and that all emission from the atom goes into the cavity.

We now consider the situation where the lasing condition
(Eq.~\ref{eq:LasingCondition}) is satisfied, and we analyze the
probability distribution of the photon number in the cavity. Deep
in the lasing regime, we can assume that the emission rate is well
approximated by $\Gamma/2$. The loss rate is still given by
Eq.~(\ref{eq:LossRate}). The peak in the photon-number probability
distribution therefore occurs at
\begin{equation}
n_{\rm max} = \frac{\Gamma}{2 \kappa}.
\label{eq:LasingAverageNumber}
\end{equation}
Note that this steady-state photon number is independent of the
atom-cavity coupling strength. It is also worth mentioning that
this relation remains valid even if $\Gamma$ is smaller than $g$.

The width of the probability distribution can be calculated as
follows. The `probability current' from the $(n-1)$-photon state
to the $n$-photon state is given by
\begin{equation}
W_{n-1 \rightarrow n} = \frac{\Gamma}{2} P_{n-1},
\end{equation}
whereas the probability current in the opposite direction is given
by
\begin{eqnarray}
W_{n \rightarrow n-1} & = & n \kappa P_{n} \nonumber
\\
& = & \frac{\Gamma}{2} P_{n} + \left( n - n_{\rm max} \right)
\kappa P_{n}.
\end{eqnarray}
Here $P_n$ is the probability of having $n$ photons in the cavity.
Using the detailed balance equation, i.e.~$W_{n-1 \rightarrow
n}=W_{n \rightarrow n-1}$, the above two equations can be combined
to give
\begin{equation}
\frac{P_n-P_{n-1}}{P_n} \approx - \, \frac{n-n_{\rm max}}{n_{\rm
max}},
\end{equation}
which can be integrated to give the probability distribution
\begin{equation}
P_n = P_{\rm max} \exp \left\{ - \, \frac{\left( n-n_{\rm max}
\right)^2}{2 n_{\rm max}} \right\}.
\end{equation}
The width of the probability distribution is therefore of the
order of $\sqrt{n_{\rm max}}$, as would be expected for the lasing
state.

We now turn to the situation where lasing is suppressed, i.e.~when
$4g^2<\Gamma\kappa$. In the linear regime (i.e.~when $n$ is
small), we can write simple detailed balance equations for the
probabilities $P_n$;
\begin{equation}
\frac{P_{n+1}}{P_n} = \frac{\Gamma_{\rm emission}(n)}{\Gamma_{\rm
loss}(n+1)} = \frac{4 g^2}{\Gamma \kappa}.
\end{equation}
This equation can be identified as the detailed-balance equation
for a cavity in thermal equilibrium at effective temperature
\begin{equation}
T_{\rm eff} = \frac{\hbar\omega_0}{k_B} \left[\log \left\{
\frac{\Gamma\kappa}{4 g^2} \right\} \right]^{-1}.
\end{equation}
Note that here we are neglecting the small ambient temperature of
the cavity. Using the Bose-distribution formula, we find that the
average number of photons at the above effective temperature is
given by
\begin{equation}
\avg{n} = \left(\frac{\Gamma\kappa}{4 g^2}-1\right)^{-1}.
\label{eq:ThermalAverageNumber}
\end{equation}
Therefore, if we start from large values of $\Gamma$ and we
gradually decrease it, the average number of photons in the cavity
starts increasing. This number follows a $1/x$-type function that
diverges at the threshold condition. The nonlinearity in the
emission rate [see Fig.~2(c)] prevents the photon number from
diverging at the critical value of $\Gamma$; instead the system
changes behaviour and enters the lasing regime.

\subsection{Semiclassical derivation}

In this subsection, we briefly review a mean-field approximation
that can be used to find an analytic expression for the number of
photons in the cavity in the lasing state. We follow closely the
calculation of Ref.~\cite{Mu} (see also Ref.~\cite{Gurvitz}): we
write equations of motion for the expectation values of the
relevant operators (in the rotating frame for simplicity), and
from the stationary steady-state solution we extract the number of
photons in the cavity.

We start with the Lindblad master equation for the model under
consideration (see e.g.~Ref.~\cite{Gardiner})
\begin{eqnarray}
\frac{d\rho}{dt} & = & - \frac{i}{\hbar} \left[ \hat{H} , \rho
\right] + \Gamma \left( \hat{\sigma}_+ \rho \hat{\sigma}_- -
\frac{1}{2} \hat{\sigma}_- \hat{\sigma}_+ \rho - \frac{1}{2} \rho
\hat{\sigma}_- \hat{\sigma}_+ \right) \nonumber \\ & & + \kappa
\left( \hat{a} \rho \hat{a}^\dagger - \frac{1}{2} \hat{a}^\dagger
\hat{a} \rho - \frac{1}{2} \rho \hat{a}^\dagger \hat{a} \right),
\label{eq:Master_equation}
\end{eqnarray}
where $\rho$ is the total system's density matrix, and
$\hat{\sigma}_{\pm}$ are the atom's raising and lowering
operators. We can now multiply this equation on the left by any
operator $\hat{A}$ and take the trace over the density matrix. The
result is an equation of motion for the average value of the
operator $\hat{A}$, denoted $\avg{A}$.

The relevant equations of motion are:
\begin{eqnarray}
\frac{d\avg{a}}{dt} & = & g \avg{\sigma_-} - \frac{\kappa}{2}
\avg{a} \nonumber
\\
\frac{d\avg{\sigma_-}}{dt} & = & g \avg{a \sigma_z} -
\frac{\Gamma}{2} \avg{\sigma_-} \nonumber
\\
\frac{d\avg{\sigma_z}}{dt} & = & - 2 g \avg{ a^{\dagger} \sigma_-
+ a \sigma_+ } + \Gamma \left( 1 - \avg{\sigma_z} \right).
\label{eq:Semiclassical_equations}
\end{eqnarray}
Using the mean-field approximation (i.e.~ setting $\avg{a
\sigma_z} = \avg{a} \avg{\sigma_z}$ etc.), choosing $\avg{a}$ to
be real and setting the left-hand sides to zero (for the
steady-state solution), we find for the average number of photons
in the cavity (using the relation $\avg{n} = \avg{a^2}$)
\begin{equation}
\avg{n} = \frac{\Gamma}{2\kappa} \left( 1 - \frac{\Gamma
\kappa}{4g^2} \right).
\label{Eq:simple_N}
\end{equation}
This expression is the mean-field approximation of the number of
photons in the cavity. It predicts that deep in the lasing regime,
i.e.~when the second term inside the parentheses can be neglected,
the number of photons will be given by $\Gamma/2\kappa$. It also
predicts that the photon number will start decreasing with
increasing $\Gamma$ and will vanish when $\Gamma\kappa/(4g^2)=1$.
Both of these results agree with the results of Sec.~II.C. Note
that the semiclassical calculation deals with average values, and
therefore is not suited to describe the thermal state (which
requires knowledge of the probability distribution of the photon
occupation number in the cavity).

\subsection{Numerical calculations}

We now solve Eq.~(\ref{eq:Master_equation}) numerically for
different values of $\Gamma$, keeping $g$ and $\kappa$ fixed. As
representative quantities that manifest the differences between
the lasing and suppressed-lasing regimes, we plot in Fig.~3 the
average photon number in the cavity $\avg{n}$ and the photon
number with maximum probability $n_{\rm max}$ as functions of the
parameter $\Gamma\kappa/(4g^2)$.

\begin{figure}[h]
\includegraphics[width=8.0cm]{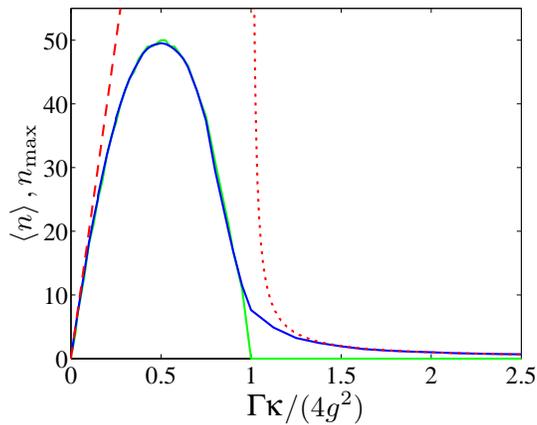}
\caption{(Color online) Average photon number $\avg{n}$ (blue
solid line) and maximum-probability photon number $n_{\rm max}$
(green solid line) in the cavity as functions of the parameter
$\Gamma\kappa/(4g^2)$. Note that $n_{\rm max}$ corresponds to red
circles in Fig.~2. The values $g/\omega_0=8\times 10^{-3}$ and
$\kappa/\omega_0=5\times 10^{-3}/(2\pi)$ were used in the
numerical calculations. The red dashed line shows the predictions
of Eq.~(\ref{eq:LasingAverageNumber}) in the lasing regime, and
the red dotted line shows the predictions of
Eq.~(\ref{eq:ThermalAverageNumber}) in the thermal regime. The
green line agrees very well with the predictions of the
semiclassical calculation (Eq.~\ref{Eq:simple_N}).}
\end{figure}

\begin{figure}[h]
\includegraphics[width=9.0cm]{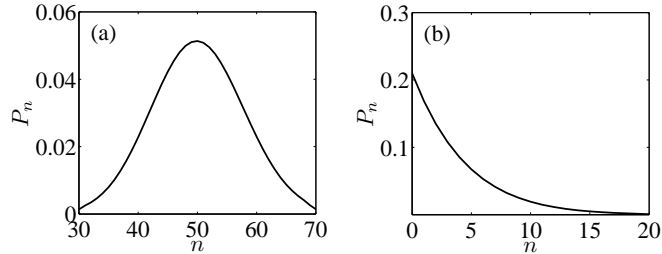}
\caption{Occupation probability as a function of photon number in
the cavity for a point in the lasing regime (a;
$\Gamma\kappa/(4g^2)=0.5$) and one in the thermal regime (b;
$\Gamma\kappa/(4g^2)=1.25$). The system parameters are given in
Fig.~3. The curve in (a) is fitted very well by a gaussian
function, and the curve in (b) is fitted very well by a Boltzmann
thermal-distribution function.}
\end{figure}

The average photon number $\avg{n}$ agrees with the analytic
expressions of Sec.~II.C [Eqs.~(\ref{eq:LasingAverageNumber}) and
(\ref{eq:ThermalAverageNumber})] away from the threshold on both
the lasing and thermal sides. The maximum-probability number
$n_{\rm max}$ agrees with the quadratic function derived in
Sec.~II.D (see also Ref.~\cite{Mu}) throughout the lasing regime.
$n_{\rm max}$ coincides with $\avg{n}$ deep in the lasing regime,
but it decreases faster as the threshold is approached and clearly
exhibits an abrupt change of behaviour when the threshold
condition is crossed.

In Fig.~4 we plot the probability distribution of the photon
number in the cavity for two points in Fig.~3, one in the lasing
state and one in the thermal state. Apart from a small regime
around the lasing-suppression threshold, the probability
distribution is fitted very well by a gaussian function in the
lasing regime and by an exponential (i.e.~Boltzmann-distribution)
function in the thermal regime.

\section{Three-level atom}

We now consider a model that corresponds more closely to the
experiment of Ref.~\cite{Astafiev}, i.e. a Cooper-pair box coupled
to a harmonic-oscillator circuit element. In the analogy with
conventional lasers using natural atoms, the Cooper-pair box plays
the role of the atom, whereas the linear circuit element plays the
role of the cavity. We follow the methods explained in Sec.~II
above and apply them to this more realistic model with a
three-level atom. It should be noted here that there has been some
work in the past on single-atom lasers using three-level atoms or
ions \cite{Mu,Briegel}. The model we consider here, however,
provides a more accurate description of the experimental situation
of main interest to us \cite{Astafiev}.

\subsection{Model}

The Hamiltonian of the system is now given by
\begin{equation}
\hat{H} = \frac{\hbar\omega_a}{2} \left( \cos \theta
\hat{\sigma}_z + \sin \theta \hat{\sigma}_x \right) +
\hbar\omega_0 \hat{a}^{\dagger} \hat{a} + g_0 \sigma_z \left(
\hat{a} + \hat{a}^{\dagger} \right),
\end{equation}
where, as before, $\omega_a$ is the atom's characteristic
frequency and $\omega_0$ is the cavity's natural frequency, and
these two frequencies are taken to be equal. The angle $\theta$
represents the deviation of the atom's bias point from the
so-called degeneracy point, and $g_0$ is the atom-cavity coupling
strength. In the three-level-atom model, the Pauli matrices
$\hat{\sigma}_x$ and $\hat{\sigma}_z$ operate on the two active
atomic states, with no need to include the third (inert) state
explicitly in the Hamiltonian (in particular, the energy of the
third state does not affect the results below).

\begin{figure}[h]
\includegraphics[width=10.0cm]{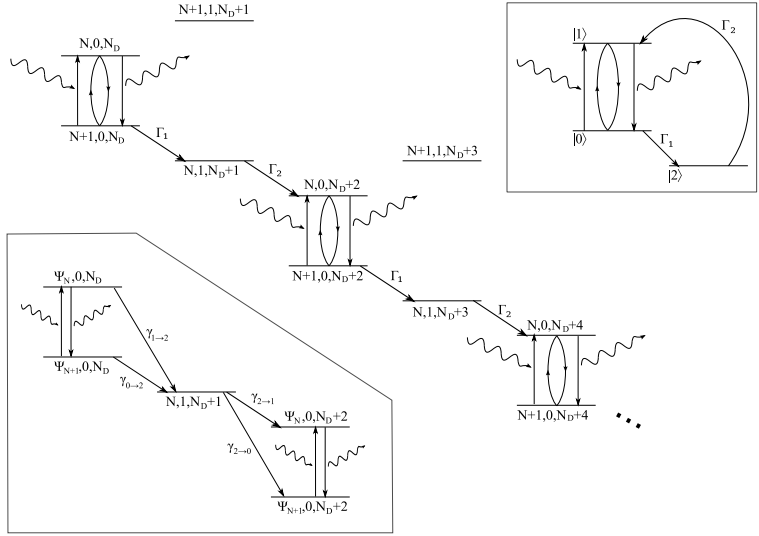}
\caption{The different processes involved in lasing for the
experiment of Ref.~\cite{Astafiev}. The first, second and third
quantum numbers represent, respectively, the number of Cooper
pairs in the box, the number of unpaired electrons in the box and
the number of electrons in the drain electrode. The box is
resonantly coupled to a cavity mode, and the two can exchange
excitations. The states with $N$ and $N+1$ Cooper pairs in the box
are coupled because the box is biased in the vicinity of the
so-called degeneracy point, so that Cooper pairs can tunnel
coherently between the box and the source electrode. The system's
total energy is lowered every time an electron tunnels (in a
dissipative process) from the box to the drain electrode. The
state with $N+1$ Cooper pairs and a single unpaired electron in
the box does not participate in the lasing mechanism. The inset in
the bottom-left corner of the figure shows the same processes as
in the main part of the figure, but in the energy eigenbasis of
the Cooper-pair box. Because of the mixing between the Cooper-pair
number states in the energy eigenstates of the box, the different
relaxation rates shown in that inset obey the relations
$\gamma_{0\rightarrow 2} = \Gamma_1 \cos^2(\theta/2)$,
$\gamma_{1\rightarrow 2} = \Gamma_1 \sin^2(\theta/2)$,
$\gamma_{2\rightarrow 1} = \Gamma_2 \cos^2(\theta/2)$ and
$\gamma_{2\rightarrow 0} = \Gamma_2 \sin^2(\theta/2)$. The inset
in the top-right corner of the figure shows a truncated model of a
three-level atom where relaxation processes take the atom from the
ground state ($\ket{0}$) to an inert state ($\ket{2}$) and then to
the excited state ($\ket{1}$).}
\end{figure}

A schematic diagram of the photon emission mechanism is shown in
Fig.~5, including the dissipative processes. The state with $N+1$
Cooper pairs in the box can relax to the state with $N$ Cooper
pairs and an unpaired electron in the box and an electron added to
the drain electrode (this transition occurs with rate $\Gamma_1$).
This state can relax further when the unpaired electron tunnels
from the box into the drain electrode (with rate $\Gamma_2$). Once
the box has $N$ Cooper pairs and no unpaired electrons, its
coupling to the source electrode allows a new Cooper pair to
tunnel from the source electrode to the box.

One can now see how the inverted relaxation process occurs. Using
an applied gate voltage to the Cooper-pair box, the system is
biased such that the state with $N+1$ Cooper pairs in the box is
lower in energy than the state with $N$ Cooper pairs, assuming a
fixed number of electrons in the drain electrode. Under this
condition, the state of the box with $N+1$ Cooper pairs can, on a
qualitative level, be identified as the ground state $\ket{0}$,
and the state with $N$ Cooper pairs in the box can be identified
as the excited state $\ket{1}$. When the box starts in the state
with $N+1$ Cooper pairs and one electron tunnels out of the box
into the drain electrode, the artificial atom goes from the ground
state $\ket{0}$ to a third state that contains $N$ Cooper pairs
and one unpaired electron in the box. Barring coincidences, the
extra unpaired electron in the box acts as an additional gate
voltage, moving the Cooper-pair box away from the degeneracy point
and from resonance with the cavity. As a result, this third level
will not be involved in any coherent dynamics and can, for our
purposes, be considered completely inert. Once in the inert state,
the unpaired electron can tunnel from the box to the drain
electrode and the atom relaxes to its excited state $\ket{1}$,
thus completing the relaxation process. The fact that the states
of the source and drain electrodes change during the relaxation
processes, which is needed in order to ensure that energy is
always lowered in each step of the relaxation process, does not
need to be explicitly taken into account once we have established
the mechanism for the inverted relaxation process in the
Cooper-pair box. Note that there are additional relaxation
processes in Fig.~5, which occur because the ground and excited
states of the box are superpositions of the states with $N$ and
$N+1$ Cooper pairs in the box.

\subsection{Photon emission and loss rates}

The derivation of the emission rate is less straightforward in
this case than the in Sec.~II.B. Nevertheless, it can be done, and
we carry it out here.

First we consider the small-$n$ limit, where photon emission can
be treated as an incoherent process. When the atom is in the state
$\ket{1}$, the process by which it emits a photon into the cavity
still occurs with the rate given by Eq.~(\ref{eq:EmissionRate}),
noting that the relevant relaxation rate here is $\Gamma_1$ (which
is the relaxation rate out of the space of active atomic states
and represents the decoherence rate in that space) and now we have
$g=g_0\sin\theta$. One difference between the present case and
that of Sec.~II.B is that now the atom can undergo recurring
transitions between the states $\ket{0}$, $\ket{1}$ and $\ket{2}$
even without the emission or absorption of photons. Therefore we
need to include some additional arguments in order to take the
above fact into account. When the atom is in the state $\ket{0}$,
the process by which it absorbs a photon from the cavity occurs
with the same rate as the one for photon emission. The net photon
emission rate is therefore given by
\begin{equation}
\Gamma_{\rm emission} = \frac{4n g^2}{\Gamma_1} \left( P_1 - P_0
\right),
\end{equation}
where $P_j$ is the occupation probability of atomic state $j$.
Using the relaxation rates shown in Fig.~5, we can derive the
probabilities of the different states:
\begin{eqnarray}
P_0 & = & \frac{\tan^2 \frac{\theta}{2}}{\tan^2 \frac{\theta}{2} +
\cot^2 \frac{\theta}{2} + \frac{\Gamma_1}{\Gamma_2}} \nonumber
\\
P_1 & = & \frac{\cot^2 \frac{\theta}{2}}{\tan^2 \frac{\theta}{2} +
\cot^2 \frac{\theta}{2} + \frac{\Gamma_1}{\Gamma_2}} \nonumber
\\
P_2 & = & \frac{\Gamma_1/\Gamma_2}{\tan^2 \frac{\theta}{2} +
\cot^2 \frac{\theta}{2} + \frac{\Gamma_1}{\Gamma_2}},
\end{eqnarray}
which gives
\begin{equation}
P_1 - P_0 = \frac{\cos\theta}{\cos^2\theta + \left( \frac{1}{2} +
\frac{\Gamma_1}{4\Gamma_2} \right) \sin^2\theta}.
\end{equation}
With this expression we find that the net emission rate in the
small-$n$ limit is given by
\begin{equation}
\Gamma_{\rm emission} = \frac{4n g^2}{\Gamma_1} \left(
\frac{\cos\theta}{\cos^2\theta + \left( \frac{1}{2} +
\frac{\Gamma_1}{4\Gamma_2} \right) \sin^2\theta} \right).
\label{Eq:EmissionRateLinear}
\end{equation}

Deep in the lasing regime, strong coupling with the cavity causes
the atom to quickly reach equal populations of the states
$\ket{0}$ and $\ket{1}$ every time it enters the space of active
states. Relaxation from the active space to the inert state
therefore occurs with rate $\Gamma_1/2$. Relaxation from the inert
state back to the active space occurs with rate $\Gamma_2$. The
atom's resetting rate (or Cooper-pair current) is therefore given
by $(\Gamma_1/2)\times\Gamma_2/(\Gamma_1/2+\Gamma_2)$, which is
the rate for a sequence of such recurrent relaxation steps. Since
the atom has two possibilities when relaxing from the inert state
to the active space [state $\ket{1}$ with probability
$\cos^2(\theta/2)$ and state $\ket{0}$ with probability
$\sin^2(\theta/2)$], the net photon emission rate will be given by
\begin{equation}
\Gamma_{\rm emission} =
\frac{(\Gamma_1/2)\times\Gamma_2}{\Gamma_1/2+\Gamma_2} \times
\cos\theta.
\label{Eq:EmissionRateSaturated}
\end{equation}
In the following subsection, we use the above rates to describe
the state of the cavity.

\subsection{Lasing condition and possible steady states}

We can now follow the arguments of Sec.~II.C with the expressions
just derived in Sec.~III.B and describe different properties of
the system.

Setting the photon emission rate in the small-$n$ limit
(Eq.~\ref{Eq:EmissionRateLinear}) equal to the photon loss rate
(Eq.~\ref{eq:LossRate}), we find the threshold condition
\begin{equation}
\frac{\Gamma_1\kappa}{4g^2} = \frac{\cos\theta}{\cos^2 \theta +
\left( \frac{1}{2} + \frac{\Gamma_1}{4\Gamma_2} \right)
\sin^2\theta}.
\label{Eq:Threshold}
\end{equation}
Deep in the lasing regime, equating the emission rate
(Eq.~\ref{Eq:EmissionRateSaturated}) to the loss rate
(Eq.~\ref{eq:LossRate}) gives an average photon number of
\begin{equation}
\avg{n} = \frac{1}{\kappa}
\frac{(\Gamma_1/2)\times\Gamma_2}{\Gamma_1/2+\Gamma_2} \cos\theta.
\label{Eq:nLasing}
\end{equation}
In the suppressed-lasing regime, we find a thermal state with
effective temperature
\begin{equation}
T_{\rm eff} = \frac{\hbar\omega_0}{k_B} \left[\log \left\{
\frac{\Gamma_1\kappa}{4 g^2} \frac{\cos^2 \theta + \left(
\frac{1}{2} + \frac{\Gamma_1}{4\Gamma_2} \right) \sin^2\theta}
{\cos\theta} \right\} \right]^{-1}.
\end{equation}
Except for the above modified expressions, the qualitative
physical description of the system remains essentially the same as
the one given in Sec.~II.C.

\subsection{Semiclassical calculation}

We now follow the semiclassical approach to derive the average
photon number in the lasing state for the three-level-atom model.
Using Eqs.~(\ref{eq:Semiclassical_equations}) as a template and
Fig.~5 as a guide for the relevant dissipative processes, we write
the equations of motion:
\begin{widetext}
\begin{eqnarray}
\frac{d\avg{a}}{dt} & = & g \avg{\sigma_-} - \frac{\kappa}{2}
\avg{a} \nonumber
\\
\frac{d\avg{\sigma_-}}{dt} & = & g \avg{a \left(P_1-P_0\right)} -
\frac{\gamma_{0\rightarrow 2}+\gamma_{1\rightarrow 2}}{2}
\avg{\sigma_-} \nonumber
\\
\frac{d(P_1-P_0)}{dt} & = & - 2 g \avg{ a^{\dagger} \sigma_- + a
\sigma_+ } + \gamma_{0\rightarrow 2} P_0 - \gamma_{1\rightarrow 2}
P_1 + \left( \gamma_{2\rightarrow 1} - \gamma_{2\rightarrow 0}
\right) \left( 1 - P_0 - P_1 \right) \nonumber
\\
\frac{d(P_1+P_0)}{dt} & = & - \gamma_{0\rightarrow 2} P_0 -
\gamma_{1\rightarrow 2} P_1 + \left( \gamma_{2\rightarrow 1} +
\gamma_{2\rightarrow 0} \right) \left( 1 - P_0 - P_1 \right),
\end{eqnarray}
where the $\gamma$s are the different relaxation rates, and
$\sigma_-$ transforms the state $\ket{1}$ into the state
$\ket{0}$. Using the relations $\gamma_{0\rightarrow 2} = \Gamma_1
\cos^2(\theta/2)$, $\gamma_{1\rightarrow 2} = \Gamma_1
\sin^2(\theta/2)$, $\gamma_{2\rightarrow 1} = \Gamma_2
\cos^2(\theta/2)$ and $\gamma_{2\rightarrow 0} = \Gamma_2
\sin^2(\theta/2)$, the steady-state solution of the above
equations gives
\begin{eqnarray}
\avg{n} & = & \frac{1}{2\kappa} \Bigg[ \frac{ \gamma_{0\rightarrow
2} - \gamma_{1\rightarrow 2} + 2 \gamma_{2\rightarrow 0} - 2
\gamma_{2\rightarrow 1} }{ \gamma_{0\rightarrow 2} +
\gamma_{1\rightarrow 2} + 2 \gamma_{2\rightarrow 0} + 2
\gamma_{2\rightarrow 1} } \left\{ \left(
\frac{\gamma_{0\rightarrow 2}^2}{4} - \frac{\gamma_{1\rightarrow
2}^2}{4} \right) \frac{\kappa}{2g^2} + \gamma_{2\rightarrow 0} +
\gamma_{2\rightarrow 1} \right\} \nonumber \\
& & \hspace{7cm} - \frac{\left(\gamma_{0\rightarrow 2} +
\gamma_{1\rightarrow 2}\right)^2}{4} \frac{\kappa}{2g^2} +
\gamma_{2\rightarrow 1} - \gamma_{2\rightarrow 0} \Bigg] \nonumber
\\
& = & \frac{1}{2\kappa} \left[
\frac{\Gamma_1-2\Gamma_2}{\Gamma_1+2\Gamma_2} \left\{
\frac{\Gamma_1^2 \kappa \cos\theta}{8g^2} + \Gamma_2 \right\}
\cos\theta - \frac{\Gamma_1^2 \kappa}{8g^2} + \Gamma_2 \cos\theta
\right] \nonumber
\\
& = & \frac{\Gamma_1}{2\kappa} \left[
\frac{1}{1+\frac{\Gamma_1}{2\Gamma_2}} \cos\theta - \left( 1 +
\frac{1-\frac{\Gamma_1}{2\Gamma_2}}{1+\frac{\Gamma_1}{2\Gamma_2}}
\cos^2\theta \right) \frac{\Gamma_1^2 \kappa}{8g^2} \right].
\label{Eq:complicated_N}
\end{eqnarray}
\end{widetext}
Equation (\ref{Eq:complicated_N}) is clearly more complicated than
Eq.~(\ref{Eq:simple_N}), reflecting the more complicated nature of
the two-step relaxation process in the three-level-atom model.

If we go deep into the lasing regime, i.e. by neglecting the terms
containing $\Gamma_1 \kappa/g^2$ in Eq.~(\ref{Eq:complicated_N}),
we recover Eq.~(\ref{Eq:nLasing}). If we equate $\avg{n}$ to zero,
we recover the threshold condition in Eq.~(\ref{Eq:Threshold}). If
we take the case where $\Gamma_1=\Gamma_2$,
Eq.~(\ref{Eq:complicated_N}) reduces to
\begin{equation}
\avg{n} = \frac{\Gamma_1}{2 \kappa} \left[ \frac{2}{3} \cos\theta
- \frac{\Gamma_1\kappa}{8g^2} \left( 1+ \frac{1}{3} \cos^2\theta
\right) \right],
\end{equation}
and the threshold condition is given by
\begin{equation}
\frac{\Gamma_1\kappa}{4g^2} = \frac{4 \cos\theta}{3+\cos^2\theta}.
\end{equation}

If we take the limit $\Gamma_2\gg\Gamma_1$,
Eq.~(\ref{Eq:complicated_N}) reduces to
\begin{equation}
\avg{n} = \frac{\Gamma_1}{2\kappa} \left[ \cos\theta -
\frac{\Gamma_1 \kappa}{8g^2} \left( 1 + \cos^2\theta \right)
\right],
\end{equation}
and the threshold condition is given by
\begin{equation}
\frac{\Gamma\kappa}{4g^2} = \frac{2\cos\theta}{1+\cos^2\theta}.
\end{equation}

For the parameters quoted in Ref.~\cite{Astafiev},
i.e.~$g=(2\pi)\times 44$ MHz, $\Gamma_1=\Gamma_2=(2\pi)\times 600$
MHz, $\kappa=(2\pi)\times 1.3$ MHz, $\theta=0.18\pi$, one finds
that the ratio $\Gamma_1\kappa/(4g^2) \approx 0.1$ (and
$\avg{n}\approx 70$), with the threshold occuring at
$\Gamma_1\kappa/(4g^2)\approx 0.9$. This set of parameters is
therefore well inside the lasing regime. By reducing $g$ and
increasing $\kappa$ (e.g.~during fabrication), however, the
boundary between the two regimes seems to be easily reachable.
Since the pumping rate $\Gamma$ is somewhat controllable in
experiment, it should be possible to study the transition between
the two regimes on a single sample.

\subsection{Numerical calculations}

We solve the quantum-optical master equation relevant to this
model [which follows straightforwardly from
Eq.~(\ref{eq:Master_equation}) and Fig.~5] numerically for
different values of $\Gamma_1$, keeping $g$, $\kappa$ and
$\Gamma_2/\Gamma_1$ fixed. We plot in Fig.~6 the average photon
number in the cavity $\avg{n}$ and the photon number with maximum
probability $n_{\rm max}$ as functions of $\Gamma_1\kappa/(4g^2)$.
The main features in Fig.~6 are similar to those in Fig.~3, which
is an indication that a good intuitive understanding of the system
can be obtained from the simplified two-level-atom model. The
curves in Fig.~6 also agree with the analytic expressions given in
this section. We do not plot the probability distributions here
because they look very similar to the ones shown in Fig.~4.

\begin{figure}[h]
\includegraphics[width=8.0cm]{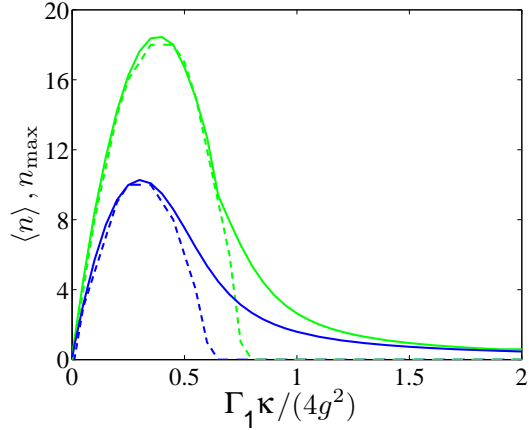}
\caption{(Color online) Average photon number $\avg{n}$ (solid
lines) and maximum-probability photon number $n_{\rm max}$ (dashed
lines) in the cavity as functions of the parameter
$\Gamma_1\kappa/(4g^2)$. The values $g/\omega_0=8\times 10^{-4}$,
$\kappa/\omega_0=5\times 10^{-4}/(2\pi)$ and $\theta=\pi/3$ were
used in the numerical calculations. The ratio $\Gamma_2/\Gamma_1$
is 1 for the blue (black) lines and 10 for the green (gray) lines.
All the numerical results agree well with theoretical predictions.
The small difference between the solid and dashed lines deep in
the lasing regime is due to the discreteness of $n_{\rm max}$.}
\end{figure}

\section{Conclusion}

We have analyzed the lasing behaviour of a single artificial atom
in a cavity, in particular in connection with recent experiments
on superconducting charge qubits. Although increased pumping
strength initially results in a larger photon population in the
cavity, increasing the pumping rate beyond a certain point starts
to suppress the number of photons in the lasing state. When the
pumping rate reaches a critical threshold value, lasing action is
completely lost and a thermal state of the cavity is formed. We
have analyzed the properties of both the lasing and
suppressed-lasing (thermal) states. We have used a
transition-rate-based approach, semiclassical calculations and
numerical simulations in our analysis, and all three methods give
consistent results. Our analysis and results are very relevant to
the experimentally achieved situation of Ref.~\cite{Astafiev},
suggesting that experimental tests of the phenomena studied in
this paper should be possible in the near future.

\begin{acknowledgments}
We would like to thank A. Abdumalikov, O. Astafiev, P. Berman, A.
Fedorov, Y. Nakamura, A. Satanin and A. Smirnov for useful
discussions. This work was supported in part by the National
Security Agency (NSA), the Laboratory for Physical Sciences (LPS),
the Army Research Office (ARO), the National Science Foundation
(NSF) grant No.~EIA-0130383, the JSPS-RFBR 06-02-91200 and
Core-to-Core (CTC) program supported by the Japan Society for
Promotion of Science (JSPS).
\end{acknowledgments}


\begin{thebibliography}{99}
\bibitem{2pi} Note that we do not use any factors of $2\pi$ with
decay rates. In other words, the occupation probability of the
atom's ground state decreases in time as $\exp\{ - \Gamma t\}$.
\bibitem{ExplainGamma1} Note that the reason for the
appearance of the relaxation rate on the right-hand side of
Eq.~(\ref{eq:EmissionRate}) is that it represents the decoherence
rate between the two atomic states. Any relevant pure dephasing
rate would also be added to the decoherence rate entering this
calculation.
\end{thebibliography}
\end{document}